\documentclass[pra,twocolumn,groupedaddress,aps,floatfix,showpacs,fleqn]{revtex4}

\usepackage{amsmath}
\usepackage{amsfonts}
\usepackage{amstext}
\usepackage{epsfig}
\usepackage{graphicx}
\usepackage{rotating}
\usepackage{draftcopy}
\usepackage{flafter}
\usepackage[below]{placeins}
\usepackage{ccaption}
\usepackage{soul}
\usepackage{wrapfig}
\usepackage{sidecap}
\usepackage{float}
\usepackage{subfig}
\usepackage{booktabs}
\usepackage{tabularx}
\usepackage{graphicx}

\begin{document}

\bibliographystyle{apsrev}

\title{Suppression of Decoherence and Disentanglement by the Exchange Interaction}

\author{Amrit De}
\author{Alex Lang}
\author{Dong Zhou}
\author{Robert Joynt}

\affiliation{Department of Physics, University of Wisconsin - Madison, WI 53706}
\date{\today }

\begin{abstract}
Entangled qubit pairs can serve as a quantum memory or as a resource for quantum communication. The utility of such pairs is measured by how long
they take to disentangle or decohere. To answer the question of whether qubit-qubit interactions can prolong entanglement, we calculate the
dissipative dynamics of a pair of qubits coupled via the exchange interaction in the presence of random telegraph noise and $1/f$ noise. We show that for
maximally entangled (Bell) states, the exchange interaction generally suppresses decoherence and disentanglement. This suppression is more apparent for random telegraph noise if the noise is non-Markovian, whereas for $1/f$ noise the exchange interaction should be comparable in magnitude to strongest noise source. The entangled singlet-triplet superposition state of 2 qubits ($\psi _{\pm }$ Bell state) can be protected by the interaction, while for the triplet-triplet state ($\phi _{\pm }$ Bell state), it is less effective. Thus the former is more suitable for encoding quantum information.
\end{abstract}
\pacs{03.67.-Pp, 03.65.-Yz, 05.40.-a}

\maketitle

\section{Introduction}
Much theoretical and experimental effort has been directed towards studying the viability of quantum information processing (QIP) in recent years due to a series of remarkable QIP algorithms \cite{Shor1997,Grover1997,Bennett1992}. Initial concerns about quantum coherence being too fragile to be useful have been partially dispelled with the discovery of quantum error-correcting codes \cite{Calderbank1996,Steane1996,Kitaev1997,Knill1998}, quantum threshold theorems\cite{Knill1998,Aharonov1997,Aliferis2006,Aharonov2006}, decoherence-free subspaces \cite{Bacon2000,Brion2007,Lidar2008} and dynamical decoupling by using optimized pulsed sequences \cite{Kofman2001,Khodjasteh2005,Uhrig2007,Lee2008}.

Qubit decoherence can be attributed to various sources and has been investigated using models such as spin-bath models \cite{Leggett1987,Wilhelm2008,Rebentrost2009}, hyperfine interaction models \cite%
{Khaetskii2002,deSousa2003,Yao.prb.2006} and phonon induced decoherence \cite{Yu2002,Golovach2004,Stano2006}. Another common source of decoherence in solid state devices are two level systems (TLS) that generate random telegraph noise (RTN), which with a wide distribution of switching rates can
give rise to $1/f$ noise \cite{Mottonen2006,Galperin2006}. In a number of recent semiconductor quantum dot(QD) experiments, RTN is observed when the potential in the dot lines up with the electrochemical potential in the reservoir, causing electrons to randomly tunnel back and forth between the dot and the reservoir \cite{Vandersypen2004,MacLean2007,Taubert2008}. This puts limitations on the performance of a QD qubit, as such a random telegraphic current can modulate the QD's orbital wavefunction which can create magnetic noise via spin-orbit coupling mechanisms. This type of noise is also known to be important in superconducting qubits \cite{Bialczak2007}, and could affect the performance of other types of qubits as well \cite{Bellomo2008,Dajka2008,Burkard2009,Zhou2010}.

This suggests that ways be invented to prolong quantum coherence and entanglement in qubit pairs in the presence of such noise sources. In this paper we present an extremely encouraging set of results -- that the exchange interaction between the qubits can be used to suppress decoherence as well as disentanglement due to RTN. This is a natural choice as the Heisenberg exchange interaction between the qubits is often used in any case to implement various gate operations such as controlled-NOT and SWAP gates \cite{DiVincenzo.nature.2000,Levy.prl.2002,Schuch2003,Fan2005}. Thus new circuit elements are not required. Typical proposals to suppress decoherence from TLS that rely on spin echo techniques \cite{Mottonen2006,Galperin2006,Gutmann2005,Bergli2007} require additional resources and system monitoring. Our proposal can be used either as an alternative or even in addition to pulsing.

Our aim in this paper is to show that for the maximally entangled Bell states, the interaction between the qubits can be used to suppress
decoherence and disentanglement. Without any loss of generality, we first analytically and numerically show this effect using a model where a single RTN source is coupled to only one of the two qubits. We then consider a model with two uncorrelated RTN sources, with each of them coupled to a qubit and show that the interaction suppresses quantum dissipation in this case as well. Our analytical results suggests that a straightforward generalization can be made to the case of multiple uncorrelated RTN sources. We subsequently show that the qubit interaction suppresses decoherence even if the qubits are coupled to a large number of uncorrelated fluctuators with a $1/f$ noise power spectrum. These results are extremely important for QIP as much of its vaunted capabilities are due to the fact that unlike the classical bit, multiple qubits exhibit quantum entanglement which allows multiple states to be addressed simultaneously\cite{Nielsen.book.2000}. Alternatively, an effective single qubit can be created from the exchange coupled singlet-triplet states, which will be less susceptible to RTN.

The exchange interaction in our work is taken to be an externally controllable parameter. Testolin $et~al.$ \cite{Testolin2009} have proposed a model to show how the two level fluctuators itself affect the exchange interaction as a function of time, while Das \cite{Das2009} has shown that the interactions can lead to periodic disentanglement and entanglement between the qubits in contact with different environments.

In general, non-Hermitian Hamiltonians are often used to describe decay processes in open quantum systems \cite{Baker1984,Dattoli1990,deSouzaDutra2005,Fleischer2005,Rotter2009}. Our calculations are done using a recently developed quasi-Hamiltonian formalism which is suitable for describing the non-unitary temporal evolution of a quantum system acted on by a classical stochastic process \cite{Cheng2008,Joynt2009,Zhou2010}. Similar approaches have been used in Refs.\cite{Saira2007,Kuopanportti2008}. In many instances, Quantum and classical noise models can yield the same solution for decoherence if the noise power spectrums have the same line-shape. For example, in \cite{Maniscalco2006,Maniscalco2008} the analytical expression obtained for a qubit's decoherence in the presence of a thermal bosonic reservior in a lossy cavity with Lorentzian broadening is exactly the same as what we obtain for classical pure dephasing noise. More recently, Saira \textit{et al}\cite{Saira2007} have shown that within the Born approximation, many fully quantum mechanical noise models can be exactly mapped onto classical stochastic noise models, including those for RTN.

This paper is organized as follows. First we describe our single fluctuator model and the quasi-Hamiltonian method in sec.\ref{sec.model}. Then in sec.\ref{sec.results}, we give results and the exact Bloch vector solutions for some important cases. Since for QIP applications the entanglement dynamics at immediate times is often the most important, in sec.\ref{sec.small} we obtain analytical results for the short time behavior for the Bell states for any arbitrary set of noise parameters. Our results are then extended to the case of two uncorrelated fluctuators in sec.\ref{sec.Hq_2F}. This treatment of two fluctuators is then subsequently extended to treat an arbitrary number of uncorrelated fluctuators in sec.\ref{sec.1byf}, where we calculate the temporal dynamics of two interacting qubits in the presence of eight RTSs with a $1/f$ distribution. Finally we present our conclusions and implications for qubit design in sec.\ref{sec.summary}.

\section{Model and Method}\label{sec.model}


It is well known that in various materials and systems sudden step like transitions occur at random intervals of time between two or more discrete voltage levels\cite{Weissman1988}. In the case of semiconductors, these random telegraphic signals (RTS) are often attributed to trapping and release of charge carriers by defect sites.  If the fuctuators are statistically independent, then the RTS can be expressed as a sum of the contributions from individual fluctuations
\begin{eqnarray}
{\bf g}'(t)=\displaystyle\sum_i{{\bf g}_i}s_i(t)
\label{RTN}
\end{eqnarray}
where, $s_i(t)$ is a RTN sequence that switches between $\pm1$ at random intervals of time, ${\bf g}_i$ is the noise vector for the $i^{th}$ fluctuator and $\left\vert \mathbf{g}\right\vert =g$ is the noise strength.  Their autocorrelation function is $\langle{s_i(t_1)s_j(t_2)}\rangle\propto\exp(-2\gamma_i|t_1-t_2|)\delta_{ij}$ and $\gamma_i$ is the switching rate of the $i^{th}$ fluctuator. Each individual fluctuator has a Lorentzian power spectrum and a broad distribution of $\gamma$ results in a $1/f$ noise power spectrum\cite{Weissman1988}. Analytical solutions are however more tractable for few RTN sources, hence in this paper we first primarily focus on the dissipative effects of a single fluctuator. We then discuss the two fluctuator case, the results of which are then extended to treat a larger number of uncorrelated fluctuators with a $1/f$ noise power spectrum.

We use the following Hamiltonian to describe the two qubit quantum system,
\begin{eqnarray}
H &=&H_{o}+H_{noise}+H_{int} \\
&=&\mathbf{B}\cdot \lbrack \mathbf{S_{1}+S_{2}}]+s(t)\mathbf{g}\cdot \mathbf{%
S_{1}}+J\mathbf{S_{1}\cdot S_{2}}  \label{H}
\end{eqnarray}
where, $\mathbf{S}_{1}=\mathbf{I\mathrm{\otimes }\boldsymbol{\sigma }_{1}}$ and $\mathbf{S}_{2}={\boldsymbol{\sigma }_{2}}\mathrm{\otimes }\mathbf{I}$ respectively represents qubits one and two, $\boldsymbol{\sigma }$ is the triad of Pauli matrices, $\mathbf{B}$ is the steady magnetic field chosen to be in the $z$ direction for all our calculations, and $J$ is the Heisenberg interaction strength. Note that in this model the RTN is only coupled to one qubit. Earlier calculations on non-interacting qubits suggest that this is sufficient to describe all the qualitative effects \cite{Zhou2010b}. The angle $\theta $, between the noise vector $\mathbf{g}$ and magnetic field $\mathbf{B}$ is called the working point of the qubit. The Hamiltonian $H$ is written for a given realization $s\left(t\right)$ of the noise. Physical quantities are calculated by
averaging over all sequences.

Since we are primarily interested in disentanglement of the qubits, it is convenient to rewrite the two qubit Hamiltonian using the maximally entangled Bell states $\psi _{\pm }=\left( |10\rangle \pm|01\rangle \right) /\sqrt{2}$ and $\phi _{\pm }=\left( |00\rangle \pm |11\rangle \right) /\sqrt{2}$ as our basis states. Eq.\ref{H} in the $\left[\phi_{-},\phi _{+},\psi _{-},\psi _{+}\right]$ basis is
\begin{equation}
H^{\prime }=\left[
\begin{array}{cc}
H_{\phi }   & H_{xy}^{\dagger } \\
H_{xy}      & H_{\psi }%
\end{array}
\right] ,  \label{H2}
\end{equation}
where $H_{\phi }=J\mathbf{I}-[B_{z}+g_{z}s(t)]\sigma _{x}$, $H_{xy}=g_{x}s(t)\mathbf{I}-ig_{y}s(t)\sigma _{x}$ and $H_{\psi}=-[J\mathbf{I}+2J\sigma_{z}+g_{z}s(t)\sigma _{x}]$.

In general, density matrices are suitable for treating open quantum systems where one has a statistical mixture of pure states. For two qubits, the time dependent density matrix can be expressed as
\begin{equation}
\rho (t)= \frac{1}{4}\left[I+\displaystyle\sum_{\left( i,j\right)\neq\left(
0,0\right) }n_{ij}(t)\sigma _{i}\otimes \sigma _{j}\right]  \label{rho-SU4}
\end{equation}
the coherent time evolution of which is governed by the von Neumann equation.
Here $I$ is the $4\times 4$ identity matrix, $\mathbf{n}(t)$ is the fifteen-component generalized Bloch vector. We have used a notation, where two indices $i$ and $j$ (where $i,j=0,x,y,z$ and $(i,j)\neq(0,0)$) are used to denote each component of the generalized Bloch vector.
Here, $\sigma _{i}\otimes\sigma_{j}=\lambda _{k}$ are the generators of $SU(4)$ and $\sigma _{0}=\mathbf{I}$ is the $2\times 2$ identity matrix.
$\left\vert\mathbf{n}\right\vert $ can be thought of as a measure of purity. For instance, $\left\vert \mathbf{n}\right\vert=0$ is the completely mixed state.


In the recently-developed quasi-Hamiltonian formalism, the dissipative temporal dynamics of an open quantum system under the influence of classical noise is calculated by transforming a random time-dependent Hamiltonian into a time-independent non-Hermitian Hamiltonian \cite{Cheng2008,Joynt2009,Zhou2010}. This is done as follows. For a given noise realization the density matrix is unitarily time evolved as $\rho (\Delta t)=U\cdot \rho (0){\cdot }U^{\dagger }$, where $U=\exp (-iH(t)\Delta t/\hbar)$. Substituting these in Eq.\ref{rho-SU4} and using the identity $Tr\left(\lambda _{i}\lambda _{j}\right) =4\delta _{ij}$ \cite{Zhou2010b}, one obtains the following temporal transfer matrix equation

\begin{equation}
\mathbf{n}(t)=\displaystyle\prod_{m=1}^{N}\mathbf{T}_{m}\cdot \mathbf{n}(0).
\label{trans_t}
\end{equation}%

\noindent where, $N=t/\Delta {t}$ and the transfer matrix elements are given by $T_{ij}=Tr\left[ U\lambda_{i}U^{\dagger }\lambda _{j}\right] /2$. As the RTN is modeled as a classical stochastic process, its dynamics is governed by the master equation, $\dot{\mathbf{W}}(t)=\mathbf{VW{\rm\it(t)}}$ \cite{VanKampen.book}, where $\mathbf{V}$ is a matrix of transition rates (such that the sum of each of its columns is zero) and $\mathbf{W}$ is the flipping probability matrix for the TLS. If the average occupation of the two states is the same (for unbiased fluctuators), then

\begin{equation}
\mathbf{W}(t)=\frac{1}{2}\left[
\begin{array}{cc}
1+e^{-2\gamma t} & 1-e^{-2\gamma t} \\
1-e^{-2\gamma t} & 1+e^{-2\gamma t}%
\end{array}.\right]
\label{W1}
\end{equation}%

\noindent Here, $\gamma $ is the switching rate for the TLF. The combined temporal dynamics of the quantum and classical TLS, averaged over all noise sequences, can be described by

\begin{equation}
\mathbf{n}(t)=\langle f|\mathbf{\Gamma }^{N}|i\rangle \mathbf{n}(0)=\langle f|\exp (-iH_{q}t)|i\rangle \mathbf{n}(0)
\label{THq}
\end{equation}

\noindent where, $|i\rangle $ and $|f\rangle$ are the initial and final state vectors for the TLS that satisfy $\mathbf{W}|i(f)\rangle =|i(f)\rangle $. For an unbiased TLS ({\it i.e.} with equal occupation probabilities), $|i\rangle =|f\rangle =[1,1]/\sqrt{2}$. Here $\mathbf{\Gamma }^{N}=\mathbf{\Gamma }_{N}\mathbf{\Gamma }_{N-1}...\mathbf{\Gamma }_{1}$ and in the small time approximation, any of the matrices

\begin{equation}
\mathbf{\Gamma }_{j}=\mathbf{W}\odot\mathbb{T}=\left[
\begin{array}{cc}
 (1-\gamma \Delta t){\bf T}_{s=+1}& \gamma \Delta t{\bf T}_{s=+1}\\
   \gamma \Delta t{\bf T}_{s=-1} & (1-\gamma \Delta t){\bf T}_{s=-1}\\
\end{array}
\right]~~~~~~~~~
\label{Gamma}
\end{equation}

\noindent Where $\odot$ denotes a Hadamard product and $\mathbb{T}$ is a square matrix, each of whose columns consists of the transfer matrices $[{\bf T}_{s=+1},{\bf T}_{s=-1}]$.  $\mathbf\Gamma$ is now a $30\times 30$ matrices in the combined generalized Bloch vector and TLS spaces. $H_{q}=\displaystyle\lim_{\Delta t\rightarrow 0}i(\mathbf{\Gamma -I})/\Delta t$ is a time-independent non-Hermitian quasi-Hamiltonian\cite{Joynt2009}. As $H_q$ is not hermitian, the time-evolution operator $\exp(-iH_qt)$ is not unitary, which makes it suitable for treating dissipative processes in open quantum systems.

There are thus two relevant time scales that determine the noise characteristics of the system.  The correlation time of the environment (which is proportional to $\gamma^{-1}$) and the time period for the noise induced Rabi-like oscillations of the qubit ( which is proportional to $g^{-1}$). If the noise correlation time of the environment is much less than the time period of the noise induced Rabi-like oscillation ($g <<\gamma,$ ), we say that the noise is Markovian. Whereas if $g>>\gamma$ we say that the noise is non-Markovian. In the former case Redfield theory can be applied to describe dissipation, while in the latter case the methods such as the quasi-Hamiltonian method is required.

\section{Results and Discussion}\label{sec.results}

\begin{figure}[tbp]
\includegraphics[width=0.85\columnwidth]{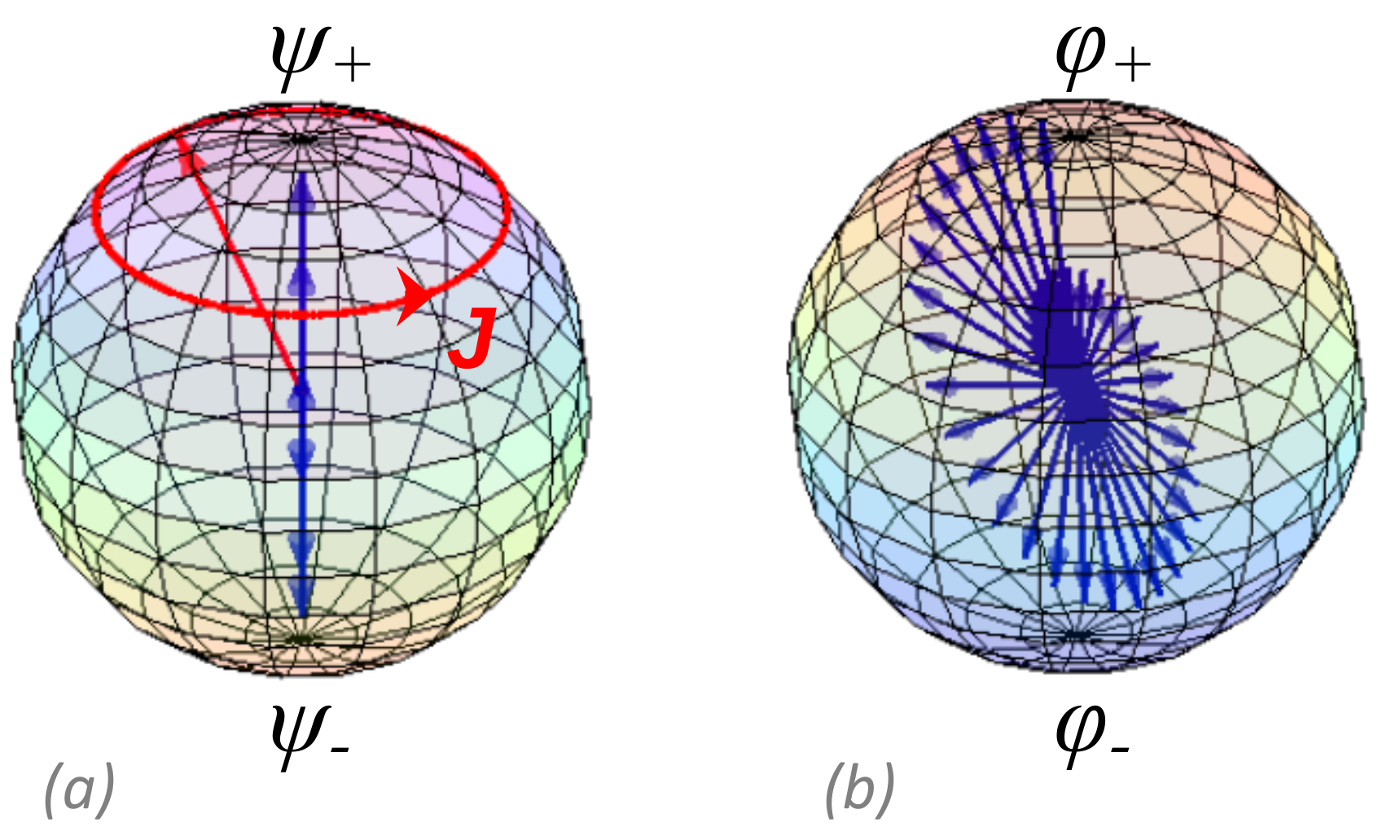}
\caption{{\protect\small(color online) Effective Bloch spheres and the ensemble averaged
Bloch vector trajectory at the pure dephasing point shown for \textbf{(a)} $%
\protect\psi _{\pm }$ and for \textbf{(b)} $\protect\phi _{\pm }$ with $%
B_{z}=1$. The free precession of $J\protect\sigma _{z}$ about $\protect\psi%
_{+}$ is also shown in \textbf{(a)}.}}
\label{BlochSphere}
\end{figure}

At the pure dephasing point ($i.e.,$ $\theta =0$), $H_{xy}=0$ in Eq.\ref{H2} and the two qubit dynamics can be effectively decoupled into two single qubit problems, whose dynamics is governed by $H_{\psi }$ and $H_{\phi }.$ The effective one particle quasi-Hamiltonian for $\psi_{\pm }$ can be extracted using the procedure outlined in the previous section.

\begin{eqnarray}
H_{q}^{\psi } &=& H_{q0}^{\psi } + H_{qJ}^{\psi}\\\notag
              &=& i\left[ \gamma (\sigma _{x}-\sigma _{0})\otimes L_{0}+g_z\sigma _{z}\otimes L_{x}\right] -i\left[2J\sigma_{0}\otimes L_{z} \right]
\label{Hq_psi}
\end{eqnarray}
where, $L_{x,y,z}\in ~SO(3)$ and $L_{0}$ is the three dimensional identity matrix. In the absence of $J$, the exact solution for the non-zero component of the Bloch vector using Eq.\ref{THq} is
\begin{equation}
n_{z}(t)=\left[ \cos(\Omega t)+\frac{\gamma }{\Omega }\sin(\Omega t)\right]e^{-\gamma t}
\label{nz_1F}
\end{equation}
where $\Omega =\sqrt{g_{z}^{2}-\gamma^{2}}$. Note that as one crosses over from the non-Markovian to the Markovian noise regime ($\gamma >g_{z}$), the trigonometric functions in $n_{z}(t)$ become hyperbolic functions and the Bloch vector's oscillations between $\psi _{+}$ and $\psi _{-}$ would then not be seen. When $g_z<<\gamma$ this reduces to the usual Redfield form $n_z(t)\approx\exp(-t/T_2)$ where, $T_2=\gamma/g^2$.

For the effect of $J$ on the Bloch vector, we approximate the matrix exponential in Eq.\ref{THq} using the Zassenhaus expansion \cite{Suzuki1977} as follows
\begin{equation}
e^{-iH_q^{\psi}t}\approx e^{-iH_{q0}^{\psi}t}e^{-iH_{qJ}^{\psi}t}e^{[H_{q0}^{\psi},H_{qJ}^{\psi}]t^2/2}
\label{Zassenhaus}
\end{equation}
This gives the following expression for the Bloch vector component, valid for either small $J,~g_z$ or at short times
\begin{equation}
n'_{z}(t)\approx\left[\zeta\cos(Jg_{z}t^{2})+\frac{g_{z}}{\Omega}\sin(\Omega{t})\sin(2Jt)\sin(Jg_{z}t^{2})\right]e^{-\gamma t}.
\label{nz_1FJ}
\end{equation}
where  $\zeta=\cos(\Omega t)+\frac{\gamma}{\Omega}\sin(\Omega t)$. Now, in case of the two qubit Bloch vector for $\psi _{\pm }$, the non zero components are $n_{xx}=n_{yy}=n_{z}$ and $n_{zz}=-1$. If in addition the initial state lies entirely in the $\psi _{\pm }$ subspace, then the entire history of the qubit pair can be visualized in the corresponding effective Bloch sphere, with $\psi _{\pm }$ at the poles (see Fig.\ref{BlochSphere}).

\begin{figure}[tbp]
\includegraphics[width=1\columnwidth]{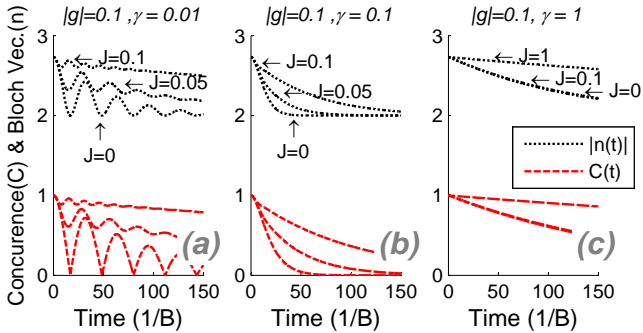}
\caption{{\protect\small(color online) Bloch vector magnitude and concurrence as a
function of time and $J$s for $\protect\psi _{+}$ shown for (a)
non-Markovian, (b) intermediate and (c) Markovian noise coupling regimes at $%
\protect\theta =0$, and $B_{z}=1$}. Note that the Bloch vector amplitudes are offset by +1 for clarity. $\gamma$, $g$ and $J$ are in units of $B$.}
\label{BVC_0}
\end{figure}

\begin{figure}[tbp]
\includegraphics[width=1\columnwidth]{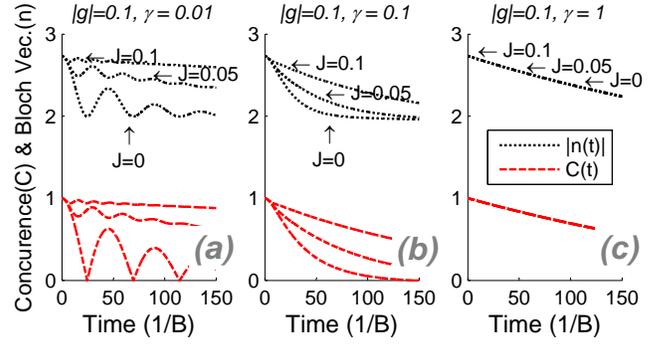}
\caption{\protect\small(color online) Same as Fig.\ref{BVC_0} but at $\theta = \pi /4$}
\label{BVC_45}
\end{figure}

Let us first consider the dynamics on the $\psi _{\pm }$ subspace, taking $\psi _{+}$ as the initial state. In the absence of $J$ in the strong
coupling limit, the effective Bloch vector oscillates between $\psi _{+}$ and $\psi _{-}$ with the quantum coherence and entanglement dissipating in time (see Figs.\ref{BlochSphere}-a and \ref{BVC_0}-a). This oscillation between $\psi _{+}$ and $\psi _{-}$ is due to the $g_{z}s(t)\sigma _{x}$ component of $H_{\psi }$. In a given noise realization, $s(t)$ causes rotations about the $x$-axis which switch randomly between the two
orientations. However the ensemble averaged Bloch vector \textit{always} travels in a straight line from pole to pole for $\psi _{\pm }$ (as shown in Fig.\ref{BlochSphere}-a) and eventually diminishes to the center, since the averaging restores the chiral symmetry. In Fig. \ref{BVC_0} we show $\left\vert \mathbf{n}\right\vert ,$ the magnitude of the Bloch vector and the concurrence as a function of time and $J$ for the $\psi _{+}$ Bell
state. For bipartite systems, the concurrence provides a measure of the entanglement between two qubits and is defined as \cite{Wootters1998}
\begin{eqnarray}
C=max\left[\sqrt\kappa_1-\sqrt\kappa_2-\sqrt\kappa_3-\sqrt\kappa_4,0\right]
\label{concur}
\end{eqnarray}
where, $\kappa _{1}$, $\kappa _{2}$, $\kappa _{3}$ and $\kappa _{4}$ are the eigenvalues of $\rho (\sigma _{y}\otimes \sigma _{y})\rho ^{\ast }(\sigma _{y}\otimes \sigma _{y})$ in decreasing order. $\left\vert \mathbf{n}\right\vert ,$ as stated above, provides a measure of the purity. The two quantities tend to track each other but are not in one-to-one correspondence. The dissipative dynamics of the entangled qubits is shown in three different noise coupling regimes --in the strong coupling limit (non-Markovian noise, $g>\gamma $), in the intermediate regime ($g\approx \gamma )$ and in the weak coupling limit (Markovian noise $g<\gamma $). Note that the purity of the state as measured by $\left\vert \vec{n}\right\vert $ is zero only at the center of the Bloch sphere whereas the concurrence is zero everywhere on the straight line that passes through the center and connects $|01\rangle$ to $|10\rangle$ (i.e. the $x$-axis).

As the exchange interaction is turned on, the $J\sigma _{z}$ component of $H_{\psi }$ causes the Bloch vector to precess about $\psi _{+}$ (see Fig.\ref{BlochSphere}-a). This effect competes with the effect of the noise, which is to drive the system toward the origin of the effective Bloch sphere. Therefore the effective ensemble averaged Bloch vector tends to remain closer to $\psi _{+}$ with increasing $J$; this delays decoherence and the disentanglement. This behavior holds true even in the intermediate noise and weak noise coupling regimes as shown in Figs.\ref{BVC_0}-b and c. However in the Markovian limit this suppression of decoherence occurs only at larger $J$ values, namely when $J\sim \gamma .$ This is shown in Fig.\ref{BVC_0}-c. The overall dissipation is also much slower for Markovian noise.

We next consider an equal mixture of dephasing and relaxational noise, a working point of $\theta =\pi /4.$. The results are shown in Fig.\ref{BVC_45}. Here no analytic solution is possible even for $J=0$ and we must diagonalize the quasi-Hamiltonian $H_{q}$ numerically. Qualitatively, the temporal behavior of dissipative process is the same as that of the pure dephasing case for all three noise coupling regimes, but with overall longer decoherence times.
\begin{figure}[tbp]
\includegraphics[width=1\columnwidth]{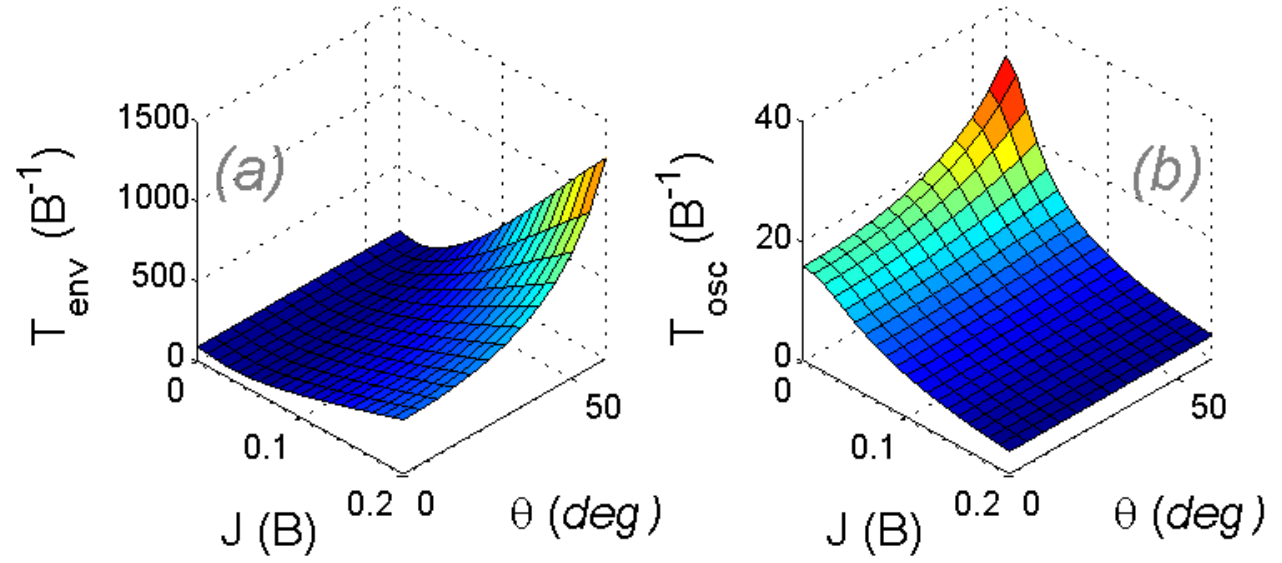}
\caption{\small(color online) Time period for the (a) envelope function decay and (b) oscillations
as a function of $J$ and $\protect\theta $. $J$ is in units of $B$.}
\label{T3T4}
\end{figure}

Clearly, the presence of the interaction $J$ suppresses disentanglement and decoherence for the $\psi _{+}$ initial state. In the non-Markovian case, however, there are several time scales, one for the envelope function decay, which fits well to a form $\sim \exp \left( -t/T_{env}\right)$ and one for the oscillation period $T_{osc}$. In Fig.\ref{T3T4}, we show $T_{env}$ and $T_{osc}$ as a function of $\theta $ and $J$.

In Fig.\ref{T3T4}-(a) it is seen that $T_{env}$ increases monotonically as a function of $\theta $. On the other hand, $T_{env}$
initially decreases as a function of $J$, reaches a minimum and then increases monotonically. This non-monotonic behavior is easily understood physically. Slow dephasing noise (small $\theta$) tends to mix $\psi _{-}$ in with the initial state $\psi _{+},$ moving the Bloch vector's trajectory "south" on through the effective Bloch sphere. Since the noise is non-Markovian, this can actually cause oscillations between $\psi_{+}$ and $\psi _{-}$. On the other hand, $J$ tends to move the trajectory "east" or "west" for a fixed realization of the noise, causing the oscillation to just miss the south pole. This results in the speeding up of the oscillations on averaging. For sufficiently large $J,$ however, the trajectory tends to stay entirely in the northern hemisphere, which slows down the oscillations. Here we have the basic mechanism for exchange interaction induced suppression of decoherence. If the initial state is an eigenstate of the interaction and the noise connects this state to a state belonging to a different eigenstate with a different eigenvalue, then the interaction term causes the trajectory to undergo tight oscillations near the initial state.

\begin{figure}[tbp]
\includegraphics[width=1\columnwidth]{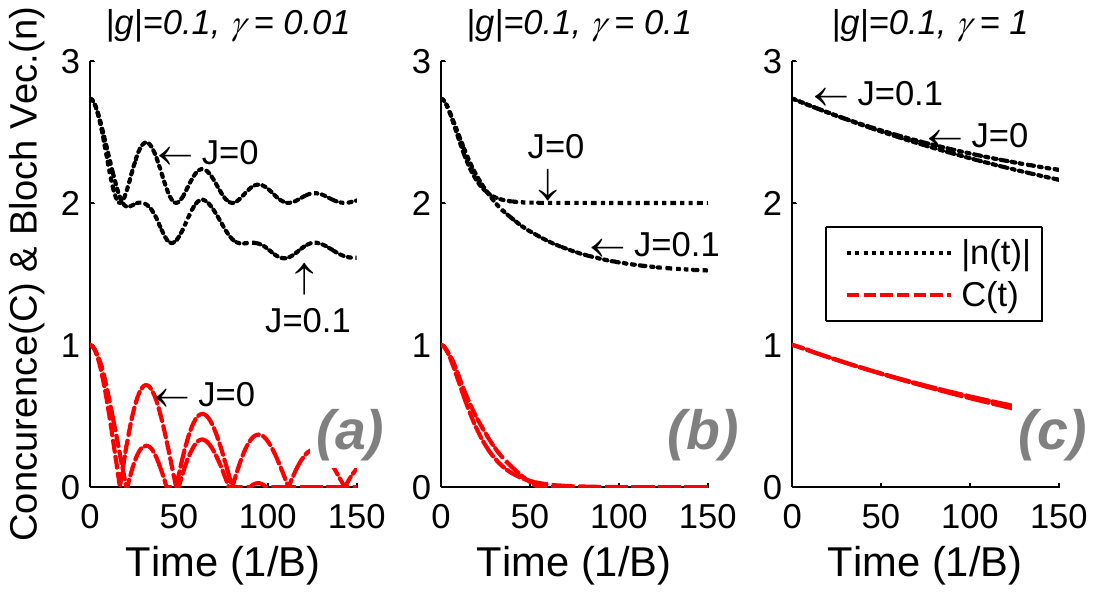}
\caption{\protect\small(color online) Bloch vector magnitude and concurence for $\protect%
\phi_+$ as a function of time and $J$s shown for (a)non-Markovian,
(b)intermediate and (c)Markovian noise coupling regimes at $\protect\theta=\protect\pi/6$ and at $B_z=10^{-3}$. The Bloch vector amplitudes are offset by +1 for clarity. $\gamma$, $g$ and $J$ are in units of $B$.}
\label{BVC_phi_30}
\end{figure}

\begin{figure}[tbp]
\includegraphics[width=1\columnwidth]{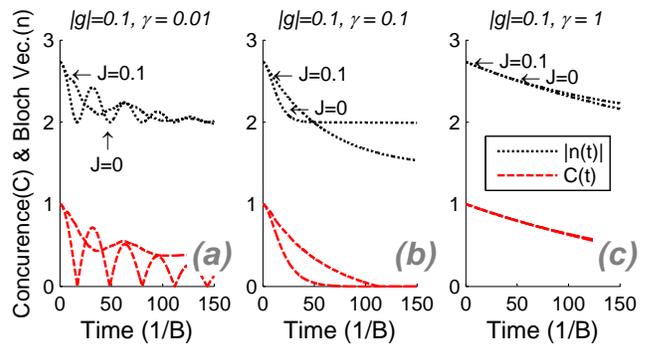}
\caption{{\protect\small(color online) Same as Fig.\protect\ref{BVC_phi_30} but at $%
\protect\theta=\protect\pi/3$.}}
\label{BVC_phi_60}
\end{figure}

To show this, we look at how the the action of $J$ depends on the initial state. Take $\phi _{+}$ as the initial state. Then the question is how this
mixes with $\phi _{-}.$ The crucial difference is that $\phi _{+}$ and $ \phi _{-}$ both belong to the triplet manifold, i.e., the exchange
interaction has the same eigenvalue for the two states.

When $\theta =0$, (pure dephasing), the qubit's dissipative dynamics is independent of $J$, as seen in $\mathbf{H}_{\phi }$ (Eq.\ref{H2}). The effective one particle quasi-Hamiltonian is therefore
\begin{equation}
H_{q}^{\phi }=i\left[ \gamma (\sigma _{x}-\sigma _{0})\otimes
L_{0}+(2B_{z}\sigma _{0}+g\sigma _{z})\otimes L_{x}\right] .  \label{Hq_phi}
\end{equation}%
The exact solution of which gives
\begin{eqnarray}
n_{y}(t) =\left[ \cos (\Omega t)+\frac{\gamma }{\Omega }\sin (\Omega t)%
\right] \cos (2B_{z}t)e^{-\gamma t} \\
n_{z}(t) =-\left[ \cos (\Omega t)+\frac{\gamma }{\Omega }\sin (\Omega t)%
\right] \sin (2B_{z}t)e^{-\gamma t}  \label{Exact_phi}
\end{eqnarray}%
The ensemble averaged temporal trajectory of the Bloch vector at a large magnetic field is shown in Fig.\ref{BlochSphere}-b. In the absence of a magnetic field, the Bloch vector only travels in a straight line from pole to pole. However, as in the case of $\psi _{\pm }$, the overall magnitude of the Bloch vector (hence the decoherence) for $\phi _{\pm }$ is independent of the magnetic field at $\theta =0$ as seen from Eq.\ref{Exact_phi}. The corresponding non-zero two qubit Bloch vector components are, $n_{xy}=n_{yx}=n_{y}$ and $n_{xx}=-n_{yy}=n_{z}$.

If $\theta \neq 0$, then $B_{z}$ suppresses the decoherence. However the effects of $J$ are not visible if $B_{z}>>J$, hence we set $B_{z}=10^{-3}$ for the next set of calculations. In Figs.\ref{BVC_phi_30} and \ref{BVC_phi_60}, the decoherence and disentanglement dynamics is shown for $\phi _{+}$ as a function of time and $J$, at various noise coupling regimes for $\theta =\pi/6$ and $\theta =\pi/3$ respectively. In the absence of $J$, the decoherence has the same qualitative behavior as that of Eq. \ref{Exact_phi}. With the onset of $J$ the decoherence is suppressed initially, however at longer times it exhibits a crossover behavior in the non-Markovian and intermediate noise coupling regimes. This crossover occurs later in time with increasing $\theta $. Thus the picture is more complicated than in the case of $\psi_\pm$ and the beneficial effect of the exchange interaction is weaker.

The results presented in this paper using the quasi-Hamiltonian method have also been confirmed through numerical simulations. A single numerical run consists of generating a sequence of random flips such that the number of
flips within a given time interval $t$ follows a Poisson distribution, $P_n(t)={(\gamma t)^n e^{-\gamma t}}/{n!}$. The time dependent density matrix is then exactly solved by numerically integrating the von Neumann equation in small steps of $\Delta t$ using the time evolution operators $U$ and $U^{\dagger}$. The final numerical result is then obtained after producing thousands of runs (5000 per dissipative curve in this paper) each with a different RTN sequence, and then averaging density matrix over all sequences. This allows us to numerically simulate the quasi-Hamiltonian results which are inherently averaged over all RTN sequences. The results from the numerical simulation and quasi-Hamiltonian method are in exact agreement to within round-off error.

\section{Analytical Results for Short Time Expansion at any Working Point}\label{sec.small}

Often tolerance levels for decoherence levels are quite stringent: a fidelity loss of more than one part in $10^{4}$ can destroy the result of a
computation \cite{Knill2005,Lucero2008}. Hence e-folding times such as $T_{1} $ and also the above $T_{env}$ are not necessarily the most relevant:it is important to look at short time behavior. We now analytically show that at short times, $J$ always suppress decoherence for the Bell states ($\phi _{\pm }$, $\psi _{\pm }$) for an arbitrary working point $\theta $, and for any set of noise parameters.

The von Neumann equation can be expanded as follows
\begin{eqnarray}
\rho(t)&=&\rho(0)-i\int_0^t[H(t'),\rho(t')]dt'\\\nonumber
       &=&\rho(0)-i\int_0^t\left[H(t_1),\rho(0)\right]dt_1\\\nonumber
       &~&+\int_0^t\int_0^{t_1}\left[H(t_2)\left[H(t_1),\rho(0)\right]\right]dt_1dt_2+...\\\nonumber
\label{rot_expand2}
\end{eqnarray}

\noindent The ensemble average of $\rho(t)$ has to be then calculated over all noise sequences
\begin{equation}
\langle{\rho}(t)\rangle= \displaystyle\sum_N(-i)^N\int_0^t\int_0^{t_1}...\int_0^{t_N}\langle{\bf C}_N\rangle dt_1dt_2...dt_N
\end{equation}
where ${\bf C}_N$ is the $N^th$ order nested commutator, ${\bf C}_N=[H(t_N),...[H(t_2), [H(t_1),\rho(0)]]]$.
The ensemble average of $\langle{\bf C}_N\rangle$ has to be then calculated over all noise sequences. This is done using the noise autocorrelation functions which are given by

\begin{eqnarray}
\langle s(t_{1})s(t_{2})...s(t_{N})\rangle &=&\langle {f}|\displaystyle%
\prod_{j=1}^{N}s(t_{j})\otimes \mathbf{W}_{j}|i\rangle \\
&=&\frac{1+(-1)^{N}}{2}\displaystyle\prod_{k=1}^{N/2}e^{-2\gamma
|t_{2k}-t_{2k-1}|}  \notag  \label{autocorrelate}
\end{eqnarray}
\noindent where $s(t_{j})=1$ for the first row of $\mathbf{W}$ and is $-1$ for the second row. $\mathbf{W}_{j}$ is as in Eq.\ref{W1} with $t\rightarrow t_{j+1}-t_{j}$. Including a factor of $(g/2)^{N}$ in the above expression accounts for the noise amplitude.

\subsection{Short time decoherence, for $\psi_\pm$ at any $\theta$}

The ensemble averaged non-zero Bloch vector components for the $\psi_\pm$ Bell states, obtained using $\rho(0)=|\psi_\pm\rangle\langle\psi_\pm|$ in the short time expansion of Eq. \ref{rot_expand2}, are listed in the Appendix along with further details. As seen in these equations, the Bloch vector components $n_{ox},n_{xo},n_{oy}$ and $n_{yo}$ depend on $J$ and do not contribute to the suppression of decoherence at pure relaxation ($\theta=\pi/2$) or pure dephasing ($\theta=0$) points. Whereas $n_{xx}$ and $n_{yy}$ depend on $J^2$. They are equal for pure dephasing and contribute to the suppression of decoherence at all working points except at pure relaxation. The Bloch vector components $n_{zx},n_{xz},n_{zy}$ and $n_{yz}$ disappear at either pure relaxation or dephasing points. These cross terms along with $n_{xy}$ and $n_{yx}$ do not contribute to the suppression of decoherence.

The following approximate expression (for short times) can be obtained for the magnitude of the $\psi_\pm$ Bloch vector by taking the sum of the squares of the above Bloch vector components
\begin{eqnarray}
\raggedleft n_{\psi}^2&\approx&3-8g_z^2\left(1-\frac{\gamma}{3}t\right)t^2 \\
&~&-\frac{2}{3}\left[12g_x^4+20g_z^2g_x^2 + g_z^2(8g_z^2+\gamma^2-J^2)\right]
t^4  \notag  \label{decohere_psi}
\end{eqnarray}
where, $n_{\psi}^2=\sum_jn_j^2$ was truncated after after the $t^4$ term. Note that the lowest order contribution from $J$ is obtained only when Eq. \ref{rot_expand2} is expanded upto fourth order in time. From the functional form of Eq.\ref{decohere_psi}, it is clearly seen that the decoherence will be suppressed with the onset of the exchange interaction.

\subsection{Short time decoherence, for $\phi_\pm$ at any $\theta$}

Similarly, in case of the $\phi_\pm$ Bell state, $\rho(0)=|\phi_\pm\rangle\langle\phi_\pm|$ is used in the short time expansion of Eq. \ref{rot_expand2}. As in the previous case, the lowest order contribution from $J$ is only obtained when Eq.\ref{rot_expand2} is expanded upto fourth order in time. The non-zero Bloch vector components are listed in the Appendix. As seen in these equations, the Bloch vector components $n_{ox},n_{xo},n_{oz}$ and $n_{zo}$ depend on $J$, however at zero magnetic field none of these components will not contribute to the suppression of decoherence in the short time expansion. $n_{oz}$ and $n_{zo}$ respectively contribute only at pure dephasing and relaxation points. Whereas $n_{ox}$ and $n_{xo}$ contribute only at intermediate working points. Note that even though all of these components depend on the sign of $J$, the overall magnitude of the Bloch vector (and hence the decoherence) is independent of the sign of $J$ (just as in for $n_\psi(t)$).

In case of the $\phi_\pm$ Bell state, the following expression for the square magnitude of the Bloch vector in the short time expansion of Eq. \ref{rot_expand2}
\begin{eqnarray}
\raggedleft n_{\phi}^2&\approx&3-8g_x^2\left(1-\frac{\gamma}{3}t\right)t^2 \\
&~&-\frac{2}{3}\left[12g_z^4+20g_x^2g_z^2 + g_x^2(8g_x^2+\gamma^2-B_z^2) %
\right]t^4  \notag  \label{decohere_phi2}
\end{eqnarray}
where, $n_{\phi}^2=\sum_jn_j^2$ was also truncated after $t^4$ term. It is seen in Eq.\ref{decohere_phi2} that unlike the case of $n_{\psi}(t)$, $J$ does not contribute in suppressing the decoherence at immediate times. Instead $B_z$ aids in maintaining the quantum coherence in its place. Even though it appears that the $J$ dependent terms will not suppress the initial decoherence due to their higher order time dependencies, our calculations show that the initial $t^2$-decoherence can be compensated for with a sufficiently strong $J$ at slightly longer times.

\section{Quasi Hamiltonian results for two fluctuators}\label{sec.Hq_2F}

In this section we will demonstrate that even in the presence of two uncorrelated fluctuators coupled to the qubits, the exchange interaction still suppresses the decoherence.

The Hamiltonian for two qubits coupled to two uncorrelated RTSs is,
{\small\begin{eqnarray}
H=  [\mathbf{B}+s_1(t)\mathbf{g}_1]\cdot\mathbf{S}_1 + [\mathbf{B}+s_2(t)\mathbf{g}_2]\cdot\mathbf{S}_2+J\mathbf{S}_1\cdot\mathbf{S}_2~~~~
\label{H2f}
\end{eqnarray}}
\noindent where, $g_k$ and $s_k(t)$ is the respective noise strength and RTN sequence for the $k^{\rm th}$ fluctuator. $\theta_k$, is angle between $\mathbf{g_k}$ and $\mathbf{B}$ and is the working point of the $k^{\rm th}$ qubit. The magnetic field is taken to be in the $z$-direction.

Here we will only obtain analytical results for the case of pure dephasing for $\psi_{\pm}$. The two qubit dynamics is then once again reduced to an effective single qubit problem governed by $H_{\psi}=2J\sigma_z+[g_{z1}s_1(t)-g_{z2}s_2(t)]\sigma_x$. The flipping probability matrix for two uncorrelated and unbiased fluctuators is $\mathbf{W}=\mathbf{W}_1\otimes\mathbf{W}_2$, where $\mathbf{W}_{1(2)}$ has the same functional form as Eq.\ref{W1} but with $\gamma$ replaced by $\gamma_{1(2)}$. Analogous to Eq.\ref{Gamma}, the combined transfer matrix for the two fluctuators and the effective qubit at the $j^{\rm th}$ instance of time is

\begin{equation}
\mathbf{\Gamma }_{j}=\mathbf{W}\odot\mathbb{T}
\label{Gamma2F}
\end{equation}

\noindent where $\mathbb{T}$ is a square matrix whose each column consists of the transfer matrices $[{\bf T}_{++},{\bf T}_{+-},{\bf T}_{-+},{\bf T}_{--}]$. Here, it is implied that ${\mathbf T}_{\pm,\pm}=\mathbf{T}_{s_1=\pm 1,s_2=\pm 1}=L_0+[JL_z\pm g_{z1}L_x \pm g_{z2}L_x]\Delta{t}$. Following the procedure outlined at the end of sec.\ref{sec.model} we obtain the following quasi Hamiltonian

\begin{eqnarray}
H_{q}^{'\psi} &=& H_{qa}^{\psi} + H_{qb}^{\psi} + H_{qi}^{\psi}
\label{Hq2f_psi}
\end{eqnarray}

where,
{\small\begin{eqnarray}
H_{qa}^{\psi}  &=& i\gamma_1(\sigma_x-\sigma_0)\otimes\sigma_0\otimes L_0 + ig_{z1}\sigma_z\otimes\sigma_0\otimes L_x~~~~~~~~\\
H_{qb}^{\psi}  &=& i\gamma_2\sigma_0\otimes(\sigma_x-\sigma_0)\otimes L_0 + ig_{z2}\sigma_0\otimes\sigma_z\otimes L_x~~~~~~~~\\
H_{qi}^{\psi}  &=& i2J\sigma_0\otimes\sigma_0\otimes L_z
\label{Hq2f_psi}
\end{eqnarray}}

$H_{qa}^{\psi},~H_{qb}^{\psi}$ and $H_{qd}^{\psi}$ commute with each other. Hence in the absence of $J$, the time dependent Bloch vector can be solved for exactly. Analogous to Eq.\ref{THq}, for two unbiased and uncorrelated fluctuators, the time dependent Bloch vector is

\begin{equation}
\mathbf{n}(t)=(\langle f_2|\otimes\langle f_1|)\exp (-iH_{q}t)(|i_1\rangle\otimes|i_2\rangle) \mathbf{n}(0)
\label{THq2}\end{equation}%
where, $|i_{1(2)}\rangle=|f_{1(2)}\rangle=[1,1]/\sqrt{2}$ are the initial and final state vectors of the two unbiased fluctuators. This results in the following non-zero component of the Bloch vector

\begin{equation}
n_{z}(t)=\displaystyle\prod_{k=1}^2\left[ \cos(\Omega_k t)+\frac{\gamma_k }{\Omega_k }\sin(\Omega_k t)\right]e^{-\gamma_kt}
\label{nz_2F}
\end{equation}
where $\Omega_k =\sqrt{g_{zk}^{2}-\gamma_k ^{2}}$.

Next, for understanding the the effect of $J$ on the Bloch vector, the matrix exponential in Eq.\ref{THq2} is calculated with the inclusion of $H_{qi}^{\psi}$ using the Zassenhaus expansion (see Eq.\ref{Zassenhaus}). We derive the following approximate expression for the dephasing of the non-zero component of the effective Bloch vector, valid only at short times or for small $J$ or for small $g_{zk}$

{\small\begin{equation}
n'_{z}(t)\approx\displaystyle\prod_{k=1}^{2}\left[\zeta_k\cos(Jg_{zk}t^{2})+\frac{g_{z,k}}{\Omega_k}\sin(\Omega{t})\sin(2Jt)\sin(Jg_{zk}t^{2})\right]e^{-\gamma_kt}.
\label{nz_2FJ}
\end{equation}}
where  $\zeta_k=\cos(\Omega_kt)+\frac{\gamma_k}{\Omega_k}\sin(\Omega_kt)$. As clearly seen from Eqs.\ref{nz_2F} and \ref{nz_2FJ}, the Bloch vectors temporal dynamics in the presence of two uncorrelated fluctuators is simply a product of the single fluctuators states shown in Eqs.\ref{nz_1F} and \ref{nz_1FJ}. This suggests that the temporal dynamics of interacting qubits in the presence of multiple uncorrelated fluctuators (which can result in $1/f$ noise) can be just as well understood in the single fluctuator picture.

So far, in this section, we have analytically shown how $J$ suppresses pure dephasing for $\psi_{\pm}$ Bell state. The most general quasi Hamiltonian for two interacting qubits in the presence of two uncorrelated fluctuators with for any arbitrary set of working point, noise parameters, field strength and for any initial state is

\begin{eqnarray}
H_q &=& H_{qy} + H_{qg} + H_{qb} + H_{qJ}
\label{Hq_2q2f}
\end{eqnarray}

\noindent where
{\small\begin{eqnarray}
H_{q\gamma} &=& i[\gamma_1(\sigma_x-\sigma_0)\otimes\sigma_0 + \gamma_2\sigma_0\otimes(\sigma_x-\sigma_0)]\otimes{L'_0}\otimes{L'_0}~~~~~~\\
H_{qg}&=&i[\vec{g_1}\cdot\vec{L'}]\otimes{L'_0}\otimes\sigma_z\otimes\sigma_0+i{L'_0}\otimes[\vec{g_2}\cdot\vec{L'}]\otimes\sigma_0\otimes\sigma_z~~~~~\\
H_{qB} &=& iB_z L'_0\otimes [L'_z\otimes{L'_0} + L'_0\otimes{L'_z} ]  \\
H_{qJ} &=& iJ L'_0\otimes[L'_x\otimes\Lambda_{zx} + \Lambda_{zx}\otimes{L'_x}+...\\\notag
&~&~L'_y\otimes\Lambda_{xz}+\Lambda_{xz}\otimes{L'_y}+L'_z\otimes\Lambda_{xy} + \Lambda_{xy}\otimes{L'_z}]~~~~~~
\label{Hq_2q2f_component}
\end{eqnarray}}.

\noindent Here, $\Lambda_{zx}=(\sigma_z+\sigma_0)\otimes\sigma_x$ , $\Lambda_{xz}=\sigma_x\otimes(\sigma_z+\sigma_0)$ and $\Lambda_{xy}=\sigma_x\otimes\sigma_x-\sigma_y\otimes\sigma_y$. $L'_{i=x,y,z}$ is the $4\times4$ form of the $SO(3)$-generators $L_i$, whose first row and first column is padded with zeros. $L'_0$ is $4\times4$ identity matrix, $\vec{g}=[g_x,g_y,g_z]$ and $\vec{L'}=[L'_x,L'_y,L'_z]$. It is important to note that the quasi Hamiltonian, $H_q$ in Eq.\ref{Hq_2q2f} is a $64\times64$ matrix which when projected down using Eq.\ref{nz_2F}, acts on the $16$ component Bloch vector instead of the $15$ component {\it generalized} Bloch vector.

\begin{figure}
\includegraphics[width=1\columnwidth]{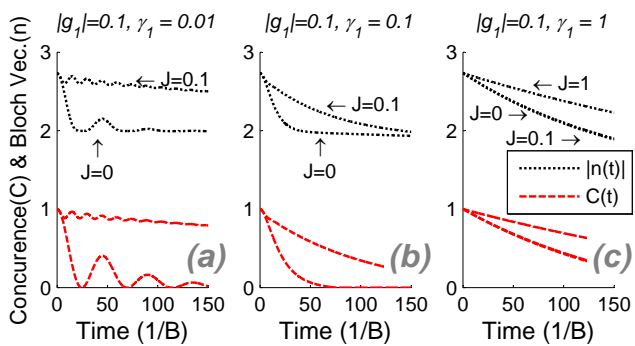}
\caption{{\protect\small(color online) Bloch vector magnitude and concurrence as a
function of time and $J$s for $\protect\psi _{+}$ in the presence of two uncorrelated RTSs shown for (a)
non-Markovian, (b) intermediate and (c) Markovian noise coupling regimes at $%
\protect\theta_1 =\pi/4$, and $B_{z}=1$}. For all calculations shown in this figure, $g_1=g_2$, $\gamma_1=\gamma_2$ and $\theta_1=\theta_2$. The the Bloch vector amplitudes are offset by +1 for clarity. $\gamma$, $g$ and $J$ are in units of $B$.}
\label{BVC2f_psi_45}
\end{figure}

The decoherence and the disentanglement dynamics for any initial Bell state can be calculated for any given set of noise parameters, by exponentiating the above quasi-Hamiltonian (Eq.\ref{Hq_2q2f}) numerically. As an example, in Fig.\ref{BVC2f_psi_45} we have shown the dissipative dynamics of the $\psi _{+}$ entangled qubits in the Markovian, intermediate and non-Markovian noise regimes for an equal mixture of dephasing and relaxational noise for both qubits $i.e.~\theta_1=\theta_2=\pi /4$. Note that if the qubit working points are held equal, then the qualitative temporal behavior of dissipative process is the same as that of pure dephasing in all three noise regimes, but with overall longer decoherence times. As seen in the figure, the decoherence and the disentanglement is significantly delayed with increasing $J$ in the non-Markovian and intermediate noise regimes. Similar to the single fluctuator case, in the Markovian limit the suppression of decoherence occurs when $J\sim\gamma$. For the sake of simplicity, all noise parameters are held equal for both fluctuators in Fig.\ref{BVC2f_psi_45}. However, $J$ will suppress the decoherence for any arbitrary choice of $\theta$s, $g$s and $\gamma$s for both $\psi_\pm$ and $\phi_\pm$ initial Bell states. This can be easily verified numerically by using Eqs.\ref{Hq_2q2f} and Eq.\ref{THq2}.

\section{Interacting Qubits in the Presence of $1/f$ Noise}\label{sec.1byf}

We now generalize our results for the interacting qubits, for a full spectrum of uncorrelated fluctuators. The general quasi Hamiltonian for two interacting qubits in the presence of $n$ uncorrelated fluctuators, for any arbitrary set of parameters, with $m$ ({\it where $m<n$}) fluctuators coupled to one qubit and $n-m$ fluctuators coupled to the second qubit is

\begin{eqnarray}
\hat{H}_q &=& \hat{H}_{qy} + \hat{H}_{qg} + \hat{H}_{qb} + \hat{H}_{qj}
\label{Hq_2qnf}
\end{eqnarray}

\noindent where
{\small\begin{eqnarray}
\hat{H}_{q\gamma} &=& i\displaystyle\sum_j\gamma_i(\tau_x^{(j)}-I_{n})\otimes{L'_0}\otimes{L'_0} ~~~\\
\hat{H}_{qg}&=&i\displaystyle\sum_{j=1}^{m}(\vec{g_j}\cdot\vec{L'})\otimes{L'_0}\otimes\tau_z^{(j)}+...\\\notag
            &~~&i\displaystyle\sum_{j=m+1}^{n}{L'_0}\otimes(\vec{g_j}\cdot\vec{L'})\otimes\tau_z^{(j)}~~~\\
\hat{H}_{qB} &=& iB_z I_{n}\otimes (L'_z\otimes{L'_0} + L'_0\otimes{L'_z} ) ~~~\\
\hat{H}_{qJ} &=& iJ I_{n}\otimes(L'_x\otimes\Lambda_{zx} + \Lambda_{zx}\otimes{L'_x}+...\\\notag
             &~&~L'_y\otimes\Lambda_{xz}+\Lambda_{xz}\otimes{L'_y}+L'_z\otimes\Lambda_{xy} + \Lambda_{xy}\otimes{L'_z}).~~~~
\label{Hq_2qnf_component}
\end{eqnarray}}

\noindent $I_n$ is an identity matrix of dimension $2^n$ and it is implied that
\begin{eqnarray}
\tau_{x(z)}^{(j)} = \sigma_0^{(1)}\otimes...\sigma_0^{(j-1)}\otimes\sigma_{x(z)}^{(j)}\otimes\sigma_0^{(j+1)}\otimes...\sigma_0^{(n)}.~~~~~
\label{tau_x}
\end{eqnarray}

\noindent The time dependent $16$ component Bloch vector is now obtained from the projected temporal dynamics of the quantum and $n$ classical TLSs as follows
\begin{equation}\small
\mathbf{n}(t)=\langle f_n|\otimes...\langle f_2|\otimes\langle f_1|\exp(-iH_{q}t)|i_1\rangle\otimes|i_2\rangle...\otimes|i_n\rangle\mathbf{n}(0).
\label{THqn}
\end{equation}

\begin{figure}
\includegraphics[width=0.95\columnwidth]{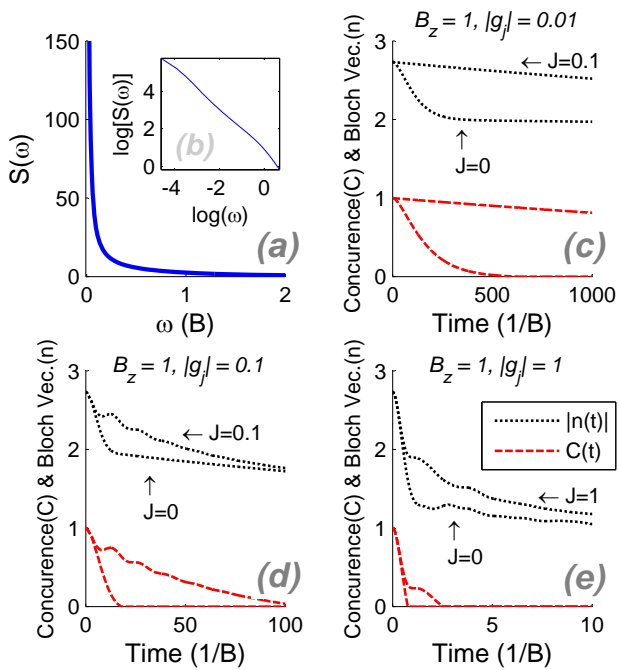}
\caption{\small(color online) (a)Power spectrum, $S(\omega)$, for a $1/f$ distribution of 8 uncorrelated fluctuators. (b) Inset showing the noise power spectrum on a logarithmic scale. Magnitude of the Bloch vector and concurrence for the $\protect\psi _{+}$ state shown as a function of time and $J$ in the presence of the same fluctuators with all $g_i$s held constant at (c) $g_j=0.01$ (d) $g_j=0.1$ (e) $g_j=1$, at $B_z=1$. $\theta=\pi/4$ for all fluctuators. The magnitude of the Bloch vector is offset by +1 for the sake of clarity and $\gamma$, $g$, $J$ are in units of $B$.}
\label{BVC1byf_psi_45}
\end{figure}

As the dimension of the quasi Hamiltonian scales as $2^n$, calculating the coupled qubit dynamics in the presence of a very large number of fluctuators can quickly become computationally very intensive. We therefore calculate the magnitude of the Bloch vector and the concurrence, using the quasi Hamiltonian in Eq.\ref{Hq_2qnf}, for a set of eight fluctuators with 4 of them coupled to each qubit. A random $1/f$ distribution of $\gamma$s is taken (ranging from $0.95$ to $0.02$), which is sufficient for generating a $1/f$ noise power spectrum (where the power spectrum is $S(\omega)=\sum\gamma_i/(\gamma_i^2+\omega^2)$) as shown in Figs. \ref{BVC1byf_psi_45}-(a) and (b). For the sake simplicity and to limit the parameter space we have held the noise strengths, $g_j$s and their respective working point the same for all the fluctuators. The coupled qubits dissipative dynamics is shown in Fig. \ref{BVC1byf_psi_45} for the $\psi_+$ state at $\theta_i=\pi/4$, for $g_j = 0.01,~0.1,~1$. In all three cases it is seen that the exchange interaction, $J$, suppresses decoherence and disentanglement. As expected the rate of decoherence rate itself increases with increasing noise strength, $g_j$.
For a full spectrum of fluctuators, if $g_j$ is smaller than the smallest $\gamma$ then the noise due to the RTSs falls purely in the Markovian noise coupling regime, as shown in Fig. \ref{BVC1byf_psi_45}-(c). If $g_j$ falls somewhere in between the selected range of $\gamma$s, then one has a mixture of Markovian and non-Markovian noise sources as shown in Fig. \ref{BVC1byf_psi_45}-(d). However in this case, the Markovian noise sources tend to dominate and the oscillatory behavior, typically seen for non-Markovian noise, tends to get washed out. Whereas for the case of $g_j=1$ in Fig. \ref{BVC1byf_psi_45}-(e), where one has a mixture of Markovian and intermediate noise sources, small oscillations superposed on top of a smoothly decaying function can be seen. This is similar to the dissipative behavior seen for a single qubit in the presence of broad spectrum noise \cite{Zhou2010}.

As in the previous case of one and two fluctuators, once the exchange interaction is turned on, the decoherence and disentanglement dynamics is suppressed for multiple fluctuators as well. This suppression is proportional to the strength of $J$. However, for the $1/f$ noise power spectrum, it is seen that the effect of $J$ in suppressing the decoherence is apparent only when it is at least the same order of magnitude as $g_j$. Finally, if all the uncorrelated fluctuators are at the pure dephasing point, then the decoherence and disentanglement dynamics of the coupled qubits are simply products of the dissipative dynamics due to each individual fluctuator. The short time behavior can be calculated for any number of fluctuators with any arbitrary distribution of $\gamma$s and $g$s by taking the $k$ sum in Eq \ref{nz_2FJ} from $1~to~n$.

\section{Summary}\label{sec.summary}

In summary we have suggested a new way to prolong quantum entanglement and coherence via qubit-qubit interaction. We have analytically and numerically shown that the exchange interaction suppresses decoherence and disentanglement for the entangled Bell states. This is shown to be true for a single RTS, for two RTSs and for $1/f$ noise as long as the fluctuators are uncorrelated. Our calculations are carried out using the quasi-Hamiltonian method, which is seen to be a particularly powerful method while dealing the many degrees of freedom associated with $1/f$ noise. For the single fluctuator case, the suppression of decoherence is most apparent when $J\sim \gamma $ and hence is more effective for non-Markovian noise. For $1/f$ noise, we conclude that $J$ should be about the same order of magnitude as the strongest noise source, $g$, in order to alleviate decoherence. As the suppression of decoherence at immediate times is key to performing high fidelity gate operations, we have analytically shown that the exchange interaction induced suppression of decoherence at short times is more effective for the $\psi _{\pm }$ Bell state then it is for the $\phi _{\pm }$ state. This is true for any arbitrary qubit working point, fluctuator switching rate and noise strength. If however a large magnetic field is used, then the $\phi _{\pm }$ Bell state is more suitable for encoding quantum information. These results are vital for quantum information processing as they can be used to develop alternative methods (in addition to existing proposals) to enhance qubit lifetimes.

\section{Acknowledgements}
This work is supported by the DARPA/MTO QuEST program through a grant from AFOSR.

$~~~~~~~~~~~~~~~~~~~~~~~~$
\section{Appendix}

In sec.\ref{sec.small}, for the $\psi_\pm$ Bell states, the ensemble averaged $\langle \rho(t)\rangle $ is obtained by using $\rho(0)=|\psi_\pm\rangle\langle\psi_\pm|$ in the short time expansion (upto $N=4$) of Eq. \ref{rot_expand2}. For any $\theta $, the lowest order contribution from $J$ is obtained only when Eq. \ref{rot_expand2} is expanded upto fourth order in time. Note that the first order term $\langle {\left[ H(t_{1}),\rho (0)\right] }\rangle =0$. Using $n_{ij}(t)=Tr\left[\sigma_{i}\otimes\sigma_{j}\times\langle\rho(t)\rangle \right]$, the following ensemble averaged non-zero Bloch vector components are obtained for $\psi_\pm$

{\small\begin{eqnarray}
n_{ox}&=& -\frac{5}{7}n_{xo} = -\frac{10}{3}B_zJg_xg_zt^4\\
n_{oy}&=& \frac{3\gamma t-4}{2-\gamma t}n_{yo} = \frac{4}{3}Jg_xg_z(2-\gamma t)t^3\\
n_{oz}&=& -n_{zo}= -\frac{2}{3}B_zJg_x^2t^4\\
n_{xy} &=&  \frac{3\gamma t-4}{\gamma t-2}n_{yx} = \frac{2}{3}B_zg_x^2(\gamma t-2)t^3 \\
n_{yz}&=& \frac{3\gamma t -4}{2-\gamma t}n_{zy} =\frac{2}{3}B_zg_xg_z( 3\gamma t - 4)t^3\\
n_{xz}&=&-\frac{2}{3}g_xg_z[3t^2-2\gamma t^3 - (3B_z^2  + 6J^2 + \Omega^2)t^4]\\
n_{zx}&=& \frac{2}{3}g_xg_z[3t^2 -\gamma t^3 + (B_z^2  + 10J^2 - \Omega^2)t^4]\\
n_{xx}&=& 1 - 2g_z^2\left(t^2  - \frac{2}{3}\gamma t^3\right)\\\nonumber
      &~&-\left[2B_z^2g_x^2 - \frac{4}{3}g_z^2(2J^2 + \Omega^2 )\right]t^4\\
n_{yy}&=& 1 - 2(g_z^2+g_x^2)\left(t^2  - \frac{2}{3}\gamma t^3\right)\\\notag
      &~& +\frac{4}{3}\left[g_x^2(\Omega^2-2Bz^2) + g_z^2(\Omega^2+2J^2)\right]t^4\\
n_{zz}&=&-1 + \frac{2}{3}g_x^2[3t^2 - 2\gamma t^3 - (B_z^2+\Omega^2)t^4]
\small\label{decohere_psi}
\end{eqnarray}}
\noindent where, $\Omega=\sqrt{g_x^2+g_z^2-\gamma^2}$ and $g_x=g_z\tan(\theta)$.

Similarly for the $\phi_\pm$ Bell state, $\rho(0)=|\phi_\pm\rangle\langle\phi_\pm|$ is used in the short time expansion of Eq. \ref{rot_expand2}. As in the case of $\psi_\pm$, Eq.\ref{rot_expand2} has to be expanded at least upto $N=4$ to see the lowest order contributions from $J$ at the pure relaxation/dephasing points. As expected, the ensemble averaged first order terms $\langle {\left[ H(t_{1}),\rho (0)\right] }\rangle =0$. The resulting non-zero Bloch vector components are

{\small\begin{eqnarray}
n_{ox}&=&-\frac{3}{5}n_{xo}=-2B_zJg_xg_zt^4 \\
n_{oz}&=&-\frac{-3g_z^2}{5g_x^2}n_{zo}=2B_zJg_z^2t^4\\
n_{xy} &=& -\frac{2}{3}B_z\left[16B_z^2 + g_x^2(2-\gamma t) + 4g_z^2(3-2\gamma t)\right]t^3~~~~~~\\
n_{yx} &=& -\frac{2}{3}B_z\left[16B_z^2 + g_x^2(8-5\gamma t) +4g_z^2(2-3\gamma t)\right]t^3~~~~~~\\
n_{yz} &=& \frac{8-6\gamma t}{20-14\gamma t}n_{zy} = \frac{8-6\gamma t}{3}B_zg_xg_zt^3\\
n_{xz} &=& \frac{2}{3}{g_xg_z}\left[3 - 2\gamma t - (3B_z^2+\Omega^2)t^2\right]t^2\\
n_{zx} &=& n_{xz} + \frac{28}{3}g_xg_zB_z^2 \\
n_{xx} &=& 1-(2g_z^2+8B_z^2)t^2 + \frac{4}{3}\gamma g_z^2t^3 \\\notag
       &~& +\frac{2}{3}\left[16B_z^4 + (5g_x^2 + 24g_z^2)B_z^2+ gz^2\Omega^2 \right]t^4\\
n_{yy} &=& g_x^2\left[2-\frac{4}{3}\gamma t -\left(\frac{14}{3}B_z^2 + \frac{2}{3}\Omega^2\right)t^2\right]t^2-n_{xx}~~~~~\\
n_{zz} &=&  1 - 2g_x^2(t^2 + \frac{2}{3}\gamma t^3 + \frac{B_z^2 + \Omega^2}{3}t^4).
\label{decohere_phi}
\end{eqnarray}}


\end{document}